\pdfoutput=1

\documentclass[pra,10pt,superscriptaddress, footnoteinbib]{revtex4}
\usepackage{amsmath}
\usepackage{latexsym}
\usepackage{amssymb}
\usepackage{graphics,epstopdf}
\usepackage[colorlinks=true, citecolor=blue, urlcolor=blue ]{hyperref}
\usepackage{epsf,graphics,graphicx}

\newcommand{\n}{\noindent}

\newcommand{\ed}{\end{document}}
\newcommand{\beq}{\begin{equation}}
\newcommand{\eeq}{\end{equation}}

\begin{document}
\title{Quantum Transport in Ferromagnetic Graphene : Role Of Berry Curvature }
\author{ Debashree Chowdhury\footnote{Electronic
address:{debashreephys@gmail.com}}${}^{}$ , Banasri Basu\footnote{Electronic address:{sribbasu@gmail.com}}${}^{}$} 
\affiliation{Physics and
Applied Mathematics Unit, Indian Statistical Institute,\\
 203
Barrackpore Trunk Road, Kolkata 700 108, India}


\begin{abstract}
\n
The magnetic effects in ferromagnetic graphene basically depend on the principle of exchange interaction when 
ferromagntism is induced by depositing an insulator layer on graphene. Here we deal with the consequences of non-uniformity in the exchange coupling strength of the ferromagnetic graphene. We  discuss how the in- homogeneity in the coordinate and momentum of  the exchange vector field can provide interesting results in the conductivity analysis of the ferromagnetic graphene. Our analysis is based on the Kubo formalism of quantum transport.

\end{abstract}

\maketitle
 
Graphene \cite{g} is a two dimensional material consisting of a single layer of carbon atoms arranged in a honeycomb or chicken wire structure. The  
familiar pencil-lead, which is known as graphite,   consists of layers of carbon atoms tightly bonded in the plane. This 
graphite layers are graphene and it is the thinnest as well as strongest material known till now. Graphene can conduct electricity as efficiently as copper and outperforms all other materials as a conductor of heat.
Graphene is almost completely transparent, yet so dense that even the smallest atom helium cannot pass through it. Unlike in ordinary semiconductors, the figure of the dispersion relation is cone like which meet at a point, the Dirac point. The
energy-momentum plot of quasiparticles behaves as if they were massless electrons, so-called Dirac fermions, that travel at a constant speed with a small but noteworthy fraction of the speed of light. Undoped graphene has a Fermi energy coinciding with the energy of the conical points.
This have completely filled valence band and empty conduction band
and there exists no bandgap in between and as such graphene is an example of gapless semiconductor
and the Hamiltonian near $K$ and $K^{'}$ points can be written as $H_K=-i\hbar v \vec{\sigma}\nabla, H_{K^{'}}=H_{K}^{T}, $
where ${\vec\sigma}$ are Pauli matrices. This form of the Hamiltonian is a two dimensional analogue of the Dirac Hamiltonian of massless fermions
but instead of $c$, we have Fermi velocity $v_{F}$( $\approx c/300 $). 
The ultra flat geometry, high electron mobility and  excellent intrinsic transport properties  make graphene a unique material in condensed matter physics. Besides, the long spin flip length of graphene makes it a promising candidate for spintronic applications.

The importance of ferromagnetism in industry and modern technology is well known. The ferromagnetism is the basis for many electrical and electro-mechanical devices, like electromagnets, electric motors, generators, transformers
and also the magnetic storag devices as tape recorders and hard disks. After the observation of graphene in isolation it was very natural for the scientists to search ferromagnetism in graphene. There are variety of ways for the experimental realization of
magnetized graphene or more precisely graphene with spin imbalance. There may exists some intrinsic ferromagnetic correlations in graphene. Use of  an insulating ferromagnetic substrate or adding a magnetic material or magnetic dopants or defect on top of the graphene sheet may be other options to achieve ferro-magnetism in graphene. In particular, by depositing ferromagnetic insulating layer (FI) on graphene, magnetization is induced through the exchange proximity interaction (EPI).  Induction of large exchange splitting has been demonstrated by depositing ferromagnetic insulator EuO on graphene \cite{za}.

On the otherhand, study of the gauge fields in spintronics \cite{zutic, fujita}, a study of the quantum mechanical spin property of carriers and its application to technology, is also a topic of recent attraction.  Although there are many paths and techniques to study the transport  of $spin$  through any solid, advantage in working with gauge fields help us to extend the usual electric and magnetic fields analysis in the richer realm of gauge fields. The Berry phase, a fundamental result in quantum mechanics, has important consequences in spintronics. The Bery phase results from cyclic, adiabatic transport of quantum states with respect to any parameter (eg. real space coordinate vector $\vec{r}$, or the momentum space vector $\vec{k}$ ). The discovery of intrinsic spin Hall effect triggered the importance of Berry phase theory in the context of spin orbit coupling. Attention was also paid to find the effects of $\vec{k}$ space gauge fields in graphene, optics and exciton systems.  

The induced magnetic effects in graphene which basically rely on the principle of exchange interaction is a topic of recent interest. In this paper,  we focus on the non-uniform exchange coupling and have studied the physics of Hall conductivity of a spin-orbit coupled (SOC)system via the Kubo formula approach. The exchange vector may be a function of coordinate, momentum or time. There is an experimental demonstration which shows  that the exchange coupling in case of the deposition of FM insulators on graphene  is momentum dependent. Moreover, space dependent exchange field is used to study the spin lens  configuration \cite{sl}. The motivation of this paper is to present the role of Berry curvature in the analysis of the Hall conductivity for the inhomogeneous exchange coupling and momentum dependent exchange coupling.

The organization of the paper is as follows: Hamiltonian of the model for the ferromagnetic graphene is developed in section I. Section II deals  with the discussion of the spin conductivity analysis for a momentum dependent exchange vector showing  the importance of the Berry curvature in this context. The next section contains the derivation of the Berry curvature for space dependent exchange coupling. Finally we summarize in the last section.
\section{The Hamiltonian}
We consider a thin insulating ferromagnetic material deposited on the top of a graphene sheet with substrate induced SOC so that the semiclassical theory of spin Hall effect in an undoped ferromagnetic graphene can be developed through the Hamiltonian \cite{za}
\begin{equation}
H = v_{F}\vec{\alpha} . \vec{k} +E_F + \vec{\sigma}.\vec{h} + V(r) + \lambda_{G} [\vec{\sigma}\times\vec{\nabla}_{r}V(r)].\vec{k},
\end{equation}
where $\alpha$ is equal to the unit matrix in spin space. $v_F$ is the Fermi velocity,  $\vec{\sigma}=(\sigma_x.\sigma_y,\sigma_z)$ is the Pauli matrix and the Fermi energy is given by $E_F$. The second term indicates the exchange Hamiltonian ($H_{ex}$) due to the interaction between the local magnetization of the ferromagnet and the surface Dirac fermions 
and $\vec{h}$ is the exchange energy vector. $V(\vec{r})$ is the total potential present in the system which includes the potential due to external electric field and crystal lattice potential $V_{crys} $. The last term denotes the spin-orbit coupling term.

In the absence of an external magnetic field, in undoped ferromagnetic graphene ($E_F=0$),  with a constant exchange energy a specific type of charge Hall effect has been predicted \cite{za}. In this case, the charge Hall effect is generated by spin Hall mechanism. Within the semiclassical theory of spin-orbital dynamics of carriers, a longitudinal electric field produces a pure charge transverse current with no polarization of spin. 

Our motivation here is to study  the spin-orbital dynamics of the carriers for the momentum dependent exchange coupling within the  semi-classical framework. We can write the Hamiltonian of the system with momentum dependent exchange field as
\begin{equation} H = v_{F}\vec{\alpha} . \vec{k}+E_F + \vec{\sigma}.\vec{h}(\vec{k}) + V(r) + \lambda_{G} [\vec{\sigma}\times\vec{\nabla}_{r}V(r)].\vec{k}\end{equation}
Collecting only the dynamical terms, we can rewrite the Hamiltonian as
\begin{equation} H(\vec{k}) = E_G({\vec k})+ E_F + \vec{\sigma}. (\vec{h}+\vec{m})(\vec{k}) = \epsilon({\vec k}) + \vec{\sigma}.\vec{M}({\vec k}),\end{equation} where $\epsilon({\vec k})= E_G({\vec k})+E_F ,$
$\vec{m}(\vec{k}) = \lambda_G\vec{\nabla}_{r}V(r)\times \vec{k}$
and $\vec{M}(\vec{k}) = (\vec{h}+ \vec{m})(\vec{k}). $  We do our analysis for exchange energy $\vec{\sigma}.\vec{h}(\vec{k})  <  $ Fermi energy $E_F,$  such that  $\alpha$ is equal to the unit matrix in the spin space and the carriers are electron like with up and down spin.

\section{Conductivity analysis}
The generalised spin orbit Hamiltonian (3) with momentum dependent exchange can now be used to derive the Hall conductivity using the Kubo method and it can be shown that the conductivity is intimately connected to the ${\vec{k}}-$ space Berry curvature.

We here use the Matsubara Greens function technique such that the conductivity can be obtained from the Kubo formula \cite{kubo} as
\begin{equation} \sigma_{xy}=  \lim_{\omega \to 0}\frac{i}{\omega}Q_{ij}(\omega + i\delta),  \end{equation}
where \begin{equation} Q_{ij} = \frac{1}{\Omega\beta}\sum_{k,n} tr\{J_{i}(\vec{k})G[k, i(\omega_{n} + \nu_{m})]J_{j}(\vec{k})G(\vec{k}, i\omega_{n})\} .\end{equation}  $G(\vec{k}, i\omega_{n})$ is the single particle Greens function, $\Omega$ is the system area, $J_{i}({\vec k)}$s are the current operator. $\omega_{n}$ and $\nu_{m}$ are the fermionic and bosonic Matsubara frequencies and can be expressed as $\omega_{n} =(2n +1)/\beta$ and $\nu_{m} = 2m\pi /\beta$. The single particle Greens function 
is given by
\begin{eqnarray}
G(\vec{k}, i\omega_{n}) &=& (i\omega_{n} - H(\vec{k}))^{-1}\nonumber\\
&=&\frac{\frac{1}{2}[1 +M_{i}(\vec{k})\sigma^{i}]/M}{i\omega_{n}-\epsilon(\vec{k}) - M} + \frac{\frac{1}{2}[1 - M_{i}(\vec{k})\sigma^{i}]/M}{i\omega_{n}-\epsilon(\vec{k}) + M}\nonumber\\
&=& \frac{R_{+}}{i\omega_{n} - E_{+}(\vec{k})} + \frac{R_{-}}{i\omega_{n} - E_{-}(\vec{k})}
\end{eqnarray}
where $M = \sqrt{M_{i}(\vec{k})M^{i}}(\vec{k}),$ $R_{\pm} = \frac{1}{2}[1 + M_{i}(\vec{k})\sigma^{i}/M]$ and $E_{\pm}(\vec{k}) = \epsilon(\vec{k}) \pm M.$ 

Using the Hamiltonian (4), the current operator $J_{i}({\vec k})$ can be derived as
\begin{equation} J_{i}(\vec{k}) = \frac{\partial H(\vec{k})}{\partial k_{i}} = \frac{\partial \epsilon(\vec{k})}{\partial k_{i}} + \frac{\partial M_{j}(\vec{k})}{\partial k_{i}}\sigma^{j},\end{equation}
where $i,j$ are the space indices. 

$Q_{ij}(i\nu_{m})$ is then given by
\begin{equation} Q_{ij}(i\nu_{m}) = \frac{1}{\Omega}\sum_{s,t=\pm}\sum_{k}\frac{tr[J_{i}(\vec{k})R_{s}(\vec{k})J_{j}(\vec{k})R_{t}(\vec{k})]}{i\nu_{m} -E_{s}(\vec{k}) + E_{t}(\vec{k})}(n_{t} - n_{s})(\vec{k}),\end{equation}
where $(n_{t} - n_{s})(\vec{k}) = (n_{F}(E_{t}(\vec{k})) - n_{F}(E_{s}(\vec{k})),$  and $n_{F}(\epsilon_{t,s}(\vec{k}))$ denotes the Fermi distribution function. 
With the help  of Matsubara sum,  the expression of $\sigma_{ij}$ can now be recast  as
\begin{equation} \sigma_{ij}=  \lim_{\omega \to 0}\frac{i}{\omega}Q_{ij}(\omega + i\delta)  .\end{equation} Inserting the value of $Q_{xy}$ in the expression we can write
\begin{eqnarray}\label{si}
\noindent \sigma_{ij} &=&  \lim_{\omega \to 0}\frac{i}{\omega}\frac{1}{\Omega}\sum_{s,t=\pm}\sum_{k}\frac{tr[J_{i}(\vec{k})R_{s}(\vec{k})J_{j}(\vec{k})R_{t}(\vec{k})]}{\omega + i\delta -E_{s}(\vec{k}) + E_{t}(\vec{k})}(n_{t} - n_{s})\nonumber\\
&=& -\frac{i}{\Omega}\sum_{k}\frac{tr[J_{i}(\vec{k})R_{-}(\vec{k})J_{j}(\vec{k})R_{+}(\vec{k}) - J_{i}(\vec{k})R_{+}(\vec{k})J_{j}(\vec{k})R_{-}(\vec{k}) ]}{(E_{+}(\vec{k}) - E_{-}(\vec{k}))^{2}}(n_{+} - n_{-})\nonumber\\
&=& -\frac{i}{\Omega}\sum_{k}\frac{tr[J_{i}(\vec{k})R_{-}(\vec{k})J_{j}(\vec{k})R_{+}(\vec{k}) -J_{i}(\vec{k})R_{+}(\vec{k})J_{j}(\vec{k})R_{-}(\vec{k}) ]}{4M^{2}}(n_{+} - n_{-})(\vec{k})\nonumber\\
& =& -\frac{i}{\Omega}\sum_{k}\frac{D(\vec{k})}{4M^{2}}(n_{+} - n_{-})(\vec{k}),
\end{eqnarray}

where $D(\vec{k}) = tr[J_{i}(\vec{k})R_{-}(\vec{k})J_{j}(\vec{k})R_{+}(\vec{k}) -J_{i}(\vec{k})R_{+}(\vec{k})J_{j}(\vec{k})R_{-}(\vec{k}) ].$
Explicitly in terms of the exchange vector, we can write
\begin{eqnarray}\label{ak}
D(\vec{k}) &=& \frac{1}{2}tr[(\frac{\partial \epsilon(\vec{k})}{\partial k_{x}} + \frac{\partial M_{\alpha}(\vec{k})}{\partial k_{x}}\sigma^{\alpha}) \frac{\frac{\partial M_{\beta}}{\partial k_{y}}M_{\gamma}}{M}(\sigma^{\beta}\sigma^{\gamma} - \sigma^{\gamma}\sigma^{\beta})]\nonumber\\
&=& \frac{2i\epsilon_{\alpha\beta\gamma}}{M}\frac{\partial M_{\alpha}}{\partial k_{x}} \frac{\partial M_{\beta}}{\partial k_{y}}M_{\gamma},
\end{eqnarray}
where $\alpha, \beta,~\gamma$ are the space indices.
Here we  have used the fact that $tr(\sigma_{\alpha}\sigma^{\beta}\sigma^{\gamma} - \sigma_{\alpha}\sigma^{\gamma}\sigma^{\beta}) = 4i\epsilon_{\alpha\beta\gamma}.$
Thus the Hall conductivity in the x-y plane can be written as
\begin{equation} \sigma_{xy} = \frac{1}{2M^{3}\Omega}\sum_{k}\epsilon_{\alpha\beta\gamma} \frac{\partial M_{\alpha}}{\partial k_{x}} \frac{\partial M_{\beta}}{\partial k_{y}}M_{\gamma}(n_{+} - n_{-})(\vec{k}).\label{34}\end{equation}
This expression shows the dependence of the Hall conductance  on the exchange field $\vec{h}(\vec{k})$ and the electric field due to the potential $V(\vec{r})$ as $\vec{M}(\vec{k}) = (\vec{h}+ \vec{m})(\vec{k}). $ 
In case of constant $\vec{h}=|h|\vec{n},$ the amplitude of the spin Hall mechanism induced charge Hall conductance is linearly proportional to exchange splitting $h$.

By converting the summation into integration over the first Brillouin zone one can write the Hall conductivity in x-y plane as \cite{kubo}
\begin{equation} \sigma_{xy} = \frac{1}{2\Omega}\int_{FBZ} \frac{c_{x}c_{y}}{4\pi^{2}}d^{2}\vec{k}\left(\vec{M}.\frac{\partial \vec{M}}{\partial k_{x}} \times\frac{\partial \vec{M}}{\partial k_{y}}\right),\end{equation}
where  $c_{x}c_{y} = \Omega.$
From the well known definition of the Berry curvature in momentum space, eqn.(13) indicates that the conductivity can be written as as
\begin{equation} \sigma_{xy} = \frac{1}{8\pi^{2}}\int_{FBZ}d^{2}k \Omega_{z}(\vec{k}),\label{con}\end{equation}
where $\Omega_{z}(\vec{k})$ is the Berry curvature in momentum space.


The Berry curvature for spin Hall systems carries a spin dependence and is equal but opposite for spin up and down electrons i.e $\Omega_{z}^{\uparrow}(\vec{k}) = - \Omega_{z}^{\downarrow}(\vec{k})$. Thus eqn (\ref{con}) indicates that
we have the spin Hall conductivity as \begin{equation} \sigma_{sH} = \sigma^{\uparrow}_{xy} - \sigma^{\downarrow}_{xy},\end{equation} whereas the charge conductivity is given by
\begin{equation} \sigma^{c}_{xy} = \sigma^{\uparrow}_{xy} + \sigma^{\downarrow}_{xy}, \end{equation}
which is effectively zero.
Thus for a undoped ferromagnetic graphene system the inhomogeneity in the exchange coupling can produce a pure transverse spin current with equal number of spin up and spin down electrons in our system.

Let us now consider the situation when the external electric potential is absent. We then have $V(\vec{r})=V_{crystal},$ only. In that case,  the Hall conductivity is given as
\begin{equation} \sigma_{xy}= \frac{1}{8\pi^{2}}\int_{FBZ} d^{2}\vec{k}\left(\vec{h}.\frac{\partial \vec{h}}{\partial k_{x}} \times\frac{\partial \vec{h}}{\partial k_{y}}\right).\end{equation}
which shows the existence of finite Hall conductivity  due to the momentum dependence of the exchange vector. 

It is amazing to note that in undoped ferromagnetic graphene, momentum dependence of the exchange vector can induce Hall conductance even in the absence of an electric field. Since the momentum dependence of exchange coupling for ferromagnetic graphene has already been demonstrated in an experiment \cite{1}, specific choice for the momentum dependence may produce some impressive result.

\section{Analysis with space dependent exchange coupling}
It is known that the space dependent exchange field can be  used to study the spin lens configuration.
For an inhomogeneous exchange vector the Hamiltonian for ferromagnetic graphene can be written as 
\begin{equation}
H = v_{F}\vec{\alpha} . \vec{k} + \vec{\sigma}.\vec{h}(\vec{r})
\end{equation}
In this case, we have not considered any external field. For the sake of simplicity the spin orbit coupling (SOC) term 
is also not considered. Actually, our motivation is to see the role of inhomogeneous exchange vector.
Variation of the exchange field can induce spin  Berry gauge as well. We can write this spin gauge as
\begin{equation} {\cal{A}}_a^{\uparrow\downarrow} = -i\left\langle \uparrow\downarrow,\vec{h}(r)|\frac{\partial}{\partial r_a}|\uparrow\downarrow,\vec{h}(r)\right\rangle = \frac{\partial h_{a}(\vec{r})}{\partial r_a}A^{\uparrow\downarrow}_{a}(h)
\end{equation}
This can also be written as 
\begin{equation}
A^{\uparrow\downarrow}_{a}(h) = \left\langle \uparrow\downarrow,\vec{h}|\frac{\partial}{\partial h_{a}}|\uparrow\downarrow,\vec{h}\right\rangle
\end{equation}
where $a=i,j,k $ are the space indices. $A^{\uparrow\downarrow}_{a}(h)$ is the exchange field dependent Berry gauge field appearing  due to the inhomogeneity of the exchange field vector. 

A physical field can be generated due to the presence of the gauge
$A^{\uparrow\downarrow}_{a}(h)$ 
It is quite obvious that the $z$ component of the curvature only exists and  is given by
\begin{eqnarray}
\Omega_{c}(\vec{r}) = \frac{\partial {\cal{A}}^{\uparrow\downarrow}_{y}}{\partial x} - \frac{\partial {\cal{A}}^{\uparrow\downarrow}_{x}}{\partial y} = \frac{\partial h_{a}}{\partial x}\frac{\partial h_{b}}{\partial y}(\frac{\partial A^{\uparrow\downarrow}_{b}}{\partial h_{a}} - \frac{\partial A^{\uparrow\downarrow}_{a}}{\partial h_{b}})
\end{eqnarray}
For further analysis, in a standard notation of unit vector we write the exchange vector as 
\begin{equation}
\vec{h} = h(sin\theta cos\phi, sin\theta sin\phi, cos\theta)
\end{equation} 
where  $\theta$ is the polar angle and $\phi$~is the azimuthal angle. 

In the adiabatic approximation, the carriers remain in the same spin eignestates for $h(r_1)$and $h(r_2),$ and the flipping  between states is forbidden. Thus we can write the adiabatic gauge as
\begin{equation}A_{ad}^{\uparrow\downarrow}(h)=\pm\frac{1}{2}(1-\cos\theta)\nabla_h\phi
\end{equation}
where  $\pm$ denotes $\uparrow$ eigenstate parallel ($\downarrow$  anti- parallel) to $h(\vec{r})$. Thus the ${\vec r}$ space Berry curvature is given by, 
\begin{equation}
\Omega_{c}(\vec{r})= \frac{\partial h_{a}}{\partial x}\frac{\partial h_{b}}{\partial y} \epsilon_{abc}(\pm \frac{h_{c}}{h^{3}}).
\end{equation}
This demonstrates  that in a ferromagnetic graphene, for the space dependent exchange vector, in the adiabatic approximation a physical field is generated 
which is the well known Berry curvature. This expression of Berry curvature shows that it is the source of monopole, if they exist.  
Further, it can be shown that in ferromagnetic graphene, if the spin orbit interaction is taken into account along with the inhomogeneous exchange vector, the Berry curvature is the origin
of the spin orbit torque \cite{future}.
\section{Summary}
Now we summarize our results. In ferromagntic graphene, conductivity depends on the inhomogeneity of the exchange vector in momentum space.
In this system momentum dependence of the exchange vector can induce Hall conductance even in the absence of an electric field. We have also shown that
this time reversal (TR) symmetric system is a potential candidate for pure spin current source. 
This study is not only mathematically interesting, it has some physical consequences as  momentum dependence of exchange coupling for ferromagnetic graphene is also experimentally verified \cite{1}. Our analysis has shown the importance of $\vec{k}-$ and $\vec{r}-$ space Berry curvature in the spin transport of ferromagnetic graphene.


\begin{thebibliography}{9}
	
	
	
	\bibitem{g} A. H. Castro Neto, F. Guinea, N. M. R. Peres
	,K. S. Novoselov and A. K. Geim, REVIEWS OF MODERN PHYSICS, {\bf 81}, (2009).
	
	\bibitem{za}Babak Zare Rameshti and Malek Zareyan, Appl. Phys. Lett. 103, 132409 (2013).
	\bibitem{zutic}I. Zutic, J. Fabian, and S. D. Sarma, Rev. Mod. Phys. {\bf 76}, 323 (2004).
	\bibitem{fujita}T Fujita, M B A Jalil and S G Tan, New Journal of Physics {\bf 12}, 013016 (2010).
	\bibitem{sl}A. G. Moghaddam and M. Zareyan, Phys. Rev. Lett. 105, 146803 (2010).
	\bibitem{kubo} R. Kubo, J. Phys. Soc. Jpn. {\bf 12}, 570, (1957).
	\bibitem{1} H. Miyazaki, T. Ito, H. J. Im, S. Yagi, M. Kato, K. Soda, and S. Kimura, Phys. Rev. Lett. {\bf 102}, 227203 (2009).
	
	
	\bibitem{future} Debashree Chowdhury and B. Basu, Annals of Physics {\bf 355}, 338,(2015).
	
\end{thebibliography}
\end{document}